\documentclass[aps,pre,twocolumn]{revtex4-1}

\usepackage{graphicx}

\usepackage{amssymb,amsfonts,amsmath,color}
\newcommand{\SetS}{\boldsymbol{\mathcal{S}}}

\begin{document}

\title{Communities as cliques}

\author{Yael Fried, David A. Kessler and Nadav M. Shnerb}

\affiliation{Department of Physics, Bar-Ilan University,
Ramat-Gan IL52900, Israel.}


\begin{abstract}
\noindent High-diversity assemblages are very common in nature, and yet the factors allowing for the maintenance of biodiversity remain obscure. The competitive exclusion principle and May's complexity-diversity puzzle both suggest that a community can support only a small number of species, turning the spotlight at the dynamics of local patches or islands, where stable and uninvadable (SU) subsets of species play a crucial role. Here we  map the community SUs question to the geometric problem of finding  maximal cliques of the corresponding graph.  We solve for the number of SUs as a function of the species richness in the regional pool, $N$, showing that this growth is subexponential, contrary to  long-standing wisdom. We show that symmetric systems relax rapidly to an SU, where the system stays until a regime shift takes place.  In asymmetric systems the relaxation time grows much faster with $N$, suggesting an excitable dynamics under noise.
\end{abstract}

\maketitle

Competition is ubiquitous in nature. Almost any aspect of living systems, from the molecular level to ecological scales, involves the competition of different species for a finite set of resources. As different populations grow, resource levels decline, putting stress on other individuals and leading to extinction of some forms of life and to saturation of others. Since Darwin this process has been recognized as the mechanism behind natural selection and evolution, meaning that competition is responsible for the apparent patterns of life on all timescales.

Still, many fundamental aspects of the theory of competition and its applicability to empirically observed patterns are far from being understood. The competitive exclusion principle \cite{gause1934struggle,hardin1960competitive}
 predicts that the maximum number of species allowed in a local community is smaller or equal to the number of limiting resources, in apparent contrast with the dozens and hundreds of species of  freshwater plankton \cite{hutchinson1961paradox,stomp2011large}, trees in tropical forests
\cite{ter2013hyperdominance} and coral reef
\cite{connolly2014commonness}. May's complexity-diversity analysis  \cite{may1972will,allesina2012stability} presents another level of difficulty; it states that, even when the number of resources is large enough, a substantial niche-overlap between species makes the chance of stable coexistence exponentially small in $N$, the species richness of the community. These long standing puzzles have received a lot of attention over the last decades, with many mechanisms  suggested to circumvent the mathematical constraints and many works that have tried to provide empirical support to these theories \cite{chesson2000mechanisms}. Nevertheless, there seems to be no general, well established and confirmed theory that explains the  persistence of high-diversity assemblages.

While some communities that support biodiversity may be considered as well-mixed, in many cases, in particular in ecology, these communities have a spatial structure. Quite generically  the dynamics takes place in local habitat patches, connected to each other by migration. Different realizations of this scenario, ranging from the  McArthur-Wilson mainland-island model (a single and relatively small patch is coupled to a well-mixed large system) \cite{losos2009theory,macarthur1967theory} to the conceptual framework of metapopulations and metacommunities (a system of many, diffusively coupled, local patches) \cite{hanski1997metapopulation,leibold2004metacommunity} have been considered in the literature.

Whichever version of these spatially structured dynamics are adopts, one immediately encounters  a fundamental problem: to identify the assembly rules of local communities \cite{diamond1975assembly, connor1979assembly} and the factors that govern their stability  \cite{law1996permanence,samuels1997divergent,chase2003community}. In the presence of strong environmental filtering one expects a one-to-one match between environment and community, rendering the effect of interspecific competition insignificant. Here we focus on the opposite scenario, where species composition is determined predominately by competition, hence  the system may support multiple steady states, and different historical sequences of species entering the local community may lead to different long-term compositions. Theory, experiments and field studies focusing on the possibility of alternative steady states (and the theory of catastrophic shifts associated with this scenario) play a central role in contemporary community dynamics literature~\cite{scheffer2001catastrophic,petraitis2013multiple}.

   To endure over intermediate/long timescales, a subset of size $S<N$ species should be intrinsically stable and uninvadable. If this subset is by itself unstable the local dynamics will drive some of the $S$ species to extinction. Even an intrinsically stable subset may still be  invaded by one of the $N-S$ species from the regional pool, rendering it a short-lived transient. A stable and uninvadable (SU) local community, on the other hand, will persist until one species goes extinct due to demographic noise or environmental variations. One then expects that the dynamics of a spatial system is dominated, on intermediate/long timescales, by SU configurations.

This insight points to a crucial question: how many SU configurations are possible, and in particular, how does this number scale with $N$?

This problem has a long history.  Gilpin and Case \cite{gilpin1976multiple} analyzed it numerically using a simple and generic description of such a community,  the generalized Lotka-Volterra equations:
\begin{equation} \label{eq1}
\frac{dx_i}{dt} = x_i - x_i \left(x_i + \sum\limits_{j \neq i }^N c_{i,j} x_j \right).
\end{equation}
Here $x_i$ is the local density of the $i$-th species population and $c_{i,j}$ is a zero-diagonal matrix of positive numbers, indicating the level of competition between the $i$-th and the $j$-th species. The larger $c_{i,j}$ is, the stronger is the stress that individuals of species $j$ put on individuals of species $i$. The SU problem, formulated for this model, goes as follows: how many size $S$-subsets of the $N$ species satisfy two conditions:

\begin{enumerate}
  \item \emph{Stability and feasibility:}  Eq. \ref{eq1}, \emph{when limited to a size $S$ subset, $\SetS$}, yields a time independent solution for which ${\bar x}_i >0$ for all of the species in $\SetS$, where ${\bar x}_i$ is the equilibrium density of the $i$-th species in the subcommunity.

  \item \emph{Uninvadability:}  Eq. \ref{eq1}, when applied to all absent $N-S$ species and linearized around the fixed point $x_i = {\bar x}_i$ for $i \in \SetS$ and $x_i = 0$ for $i \not\in \SetS$, yields negative growth rates ${\dot x}_i/x_i$ for all $i \not\in \SetS$.
\end{enumerate}

Based on their (quite limited, by today's standards) numerical simulations,  Gilpin and Case concluded that the number of SU states grows exponentially with the richness of the regional community $N$.

Fisher and Mehta~\cite{fisher2014transition}, in a new work, touched on the same problem from another perspective, the niche-neutral debate and the intermediate models suggested to bridge between these opposing approaches \cite{gravel2006reconciling,kadmon2007integrating}. They have heuristically mapped a variant of the competing species model to a known physical model for glassy behavior, the random energy model~\cite{derrida1981random}. Armed with this mapping, Fisher and Mehta interpreted the freezing transition, associated with the temperature below which a liquid shows glassy features, as the transition between niche-like and neutral-like ecological dynamics. The weak noise regime of the Lotka-Volterra dynamics was argued to correspond to the glassy phase, where local energy minima (in physical glasses) or SU states (in ecosystems) govern the dynamics, and the system spends most of its time close to one of these attractive states until it is kicked, stochastically, to another domain of attraction. Under strong noise, on the other hand, the SU/local minima structure would be washed away by the noise, rendering a neutral-like behavior.

\section{The geometry of competition}

Both the Gilpin-Case and  Fisher-Mehta works suffer from a major technical obstacle: given the interaction matrix $c_{i,j}$, it is quite difficult to find all its SU states. To do that for a system with $N$ species, one should scan through all the $2^N$ possible configurations, checking for stability/feasability and uninvadablity for each of these combinations. Therefore,  Gilpin and Case considered only limited regional pools with  $N \leq 14$, while Fisher and Mehta extracted analytic results only for their simplified, presence-absence model, that was mapped to the random energy model using strong (and not necessarily realistic) assumptions regarding the connection between invasion rates and the strength of stochasticity. To proceed, we suggest a geometric reduction of the problem.

 We consider first the symmetric version of the model. In this version, studied by both \cite{gilpin1976multiple} and \cite{fisher2014transition}, $c_{i,j} = c_{j,i}$. This symmetric scenario provides more transparent results, and we will use this case to clarify the general method.

As explained in \cite{fisher2014transition,kessler2015generalized}, the interaction matrix $c_{i,j}$ may be characterized by two parameters: the average value of an element $\overline{c}$ (overline denotes an average over all the $N(N-1)$ nondiagonal entries), reflecting the average pressure one species extracts on the other,  and the variance, $\sigma^2 = \overline {c^2}-\overline{c}^2$.

The main insight in this work is the realization that the problem of finding the SUs simplifies tremendously in
    the extreme case where the competition matrix elements are either zero or infinity. In a symmetric model this condition implies  that every pair of species, $i$ and $j$, are either mutually exclusive ($c_{i,j} = c_{j,i} = \infty$, they cannot live together in the same local community, and none of them can invade a subcommunity that includes its opponent~\cite{diamond1975assembly}) or non-interfering ($c_{i,j} = c_{j,i}=0$), meaning that they can live happily together, like the different species in the McArthur-Wilson model~\cite{losos2009theory,macarthur1967theory}.

\begin{figure*}
 \includegraphics[scale=0.5]{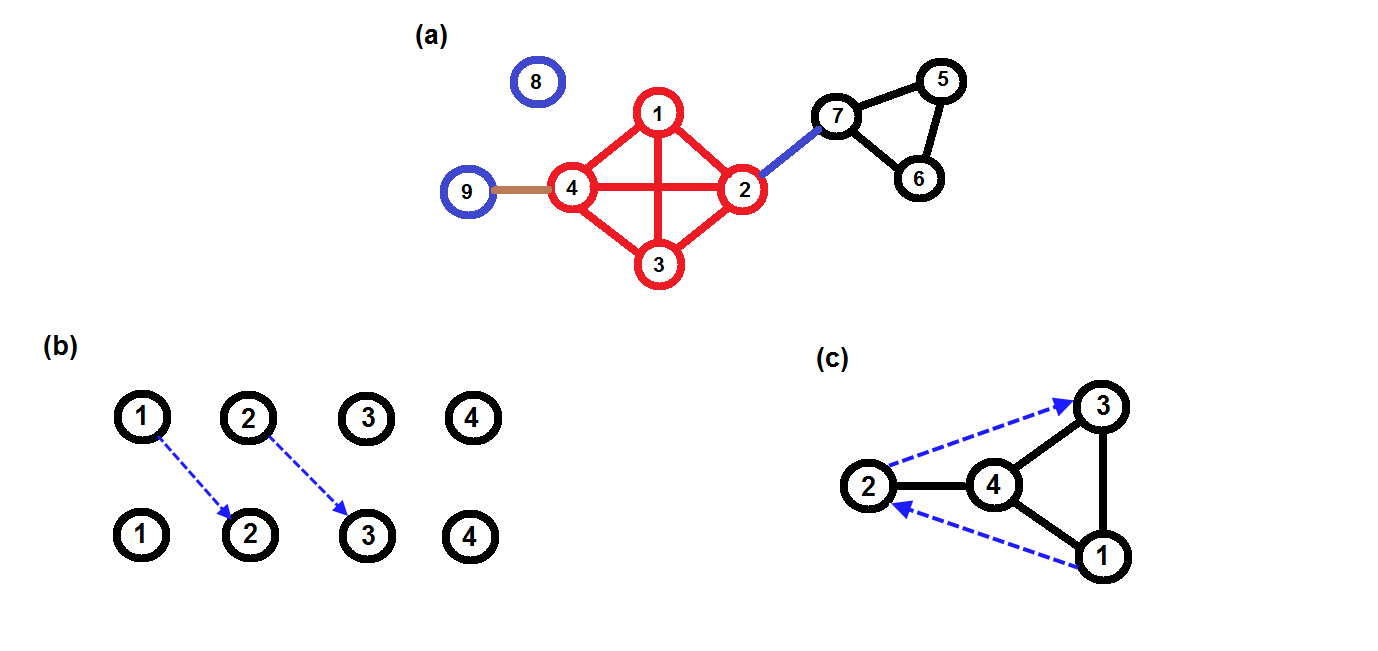}
\caption{The geometric interpretation of competition networks. In (a) an example of a network for the symmetric model is presented. Every pair of non-interacting species is connected by a full line; for example, species 1,2,3 and 4 are all non-interacting, meaning that $c_{1,2}=c_{1,3}=c_{1,4}=c_{2,4} = 0$. A clique, like $\{1,2\}$ or $\{5,6,7\}$, is a noninteracting subset of the species. A clique is stable only if another species  cannot invade it, so $\{1,2\}$ is unstable, since it may be invaded by $3$ and $4$. The stable and uninvadable subsets are the maximal cliques $\{1,2,3,4\}$, $\{5,6,7\}$,$\{4,9\}$,$\{2,7\}$ and$\{8\}$. (b) provides an example of an asymmetric system, where a dashed line represent dominance relationships. Here species 1 dominates 2 ($c_{1,2} = 0, c_{2,1} = \infty$) and species 2 dominates 3. In (c) we present this system as a network, where full lines indicate, as before, no interaction and dashed lines with arrows indicate dominance, the arrow pointing towards the inferior species. Although $\{1,3,4\}$ and $\{2,4\}$ are both maximal cliques, only  $\{1,3,4\}$ is SU (2 cannot invade since 1 dominates it) while $\{2,4\}$ is invadable by 1. }\label{fig1}
\end{figure*}

In Figure \ref{fig1} we show how to represent such a scenario by a network. First, every species is represented by a vertex, and the  $i$ and $j$
 vertices are linked (adjacent) iff the corresponding species are non-interfering. Accordingly, the $i,j$ element of the adjacency matrix of this undirected graph is unity if $c_{i,j}=0$ and zero if $c_{i,j}=\infty$.

Given this representation, the one to one correspondence between SU subcommunities and \emph{maximal cliques} of the corresponding graph is easily recognized. A clique is a collection of vertices that are all connected, one to each of the others, here representing a group of species that can all coexist.  To be uninvadable this clique should be maximal, i.e., a clique that cannot be extended by including any other adjacent vertex. Although the problem of finding all maximal cliques of a graph is known to be NP-complete~\cite{karp1972reducibility}, in practice there are efficient algorithms~\cite{bron1973algorithm} that allow one to find cliques quite easily for $N$ up to $500$, way above the numbers that have been previously considered.

 More importantly, this simple geometric interpretation of the problem allows us  to also obtain analytic results.  The expected number of maximal cliques with exactly $S$ species, $SU(N,S)$,  may be determined by multiplying  the number of subsets of size $S$ (binomial factor) by the chance that a single, randomly chosen subset is indeed a maximal clique. For random symmetric interactions the adjacency graph is an  Erd\"{o}s-Renyi network  of size $N$, and  the result is given by \cite{bollobas1976cliques},
\begin{equation}\label{eq2}
SU(N,S) = \binom{N}{S} p^{\frac{S(S-1)}{2}} (1-p^S)^{N-S},
\end{equation}
where $p$ is the probability that two randomly chosen vertices are connected.  In section I of the supplementary information (SI) we show how to find an asymptotic expression for this sum using Laplace's method,
\begin{equation}\label{eq3}
SU(N) \sim N^{a(p)\ln(N)} = e^{a(p) \ln^2(N)}
\end{equation}
where $a(p) = 1/[2\ln(1/p)]$.

\begin{figure*}
\begin{center}
\includegraphics[width=9cm]{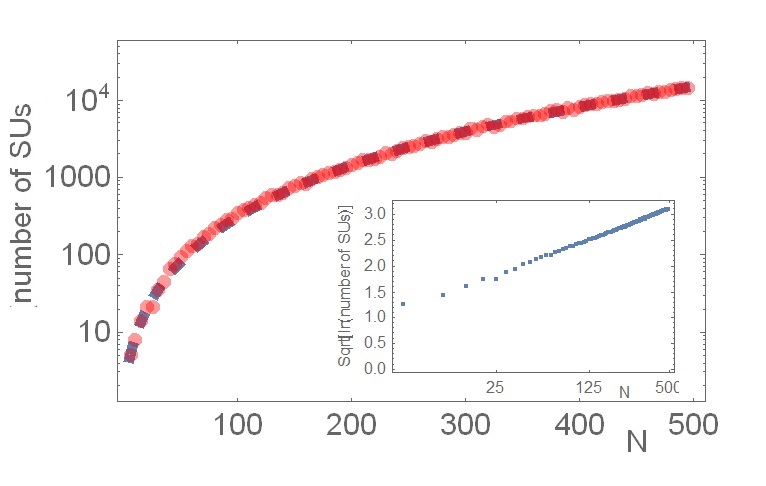}
\end{center}
  \caption{Number of maximal cliques as a function of $N$ for the symmetric zero-infinity model, plotted on semilog-y scale. Results were obtained from a  symmetric model with $p=0.1$, $N$ running from 5  to 500 in intervals of 5. Points correspond to the number of maximal cliques in a single realization, dashed line is $N^{0.22\log(N)}$. In the inset we plot $\sqrt{ln(SU)}$ on semilog-x scale, emphasizing that this is a straight line, in agreement with Eq. \ref{eq3}.
  }\label{fig2}
\end{figure*}

Figure \ref{fig2} shows the increase of $SU$ with $N$, with perfect agreement with Eq. (\ref{eq3}).  This by itself is a counter-example to the central results of \cite{gilpin1976multiple} and to the main outcome of \cite{fisher2014transition}: the number of SU states does \emph{not} increase exponentially with $N$. Therefore, the analogy that \cite{fisher2014transition} tried to establish between the random energy model, and the problem at hand, turns out to be invalid. A correct thermodynamical analog of the community dynamics problem, if there is one, would have to be a variant of the random energy model with $N^{a\ln(N)}$ number of states, as opposed to $2^N$ in its standard spin  version. In such a case the temperature of the glass transition (as before, this is analogous to the level of noise below which the system is niche-dominated, and above which the dynamics appears to be neutral) diverges like $N/\ln^2(N)$ at large $N$, meaning that there is no true neutral phase, and the system is niche dominated even if the noise is relatively large (see discussion section).

Now let us consider the more general case, where the interaction matrix has no symmetry properties but the $c_{i,j}$s are still either infinite or zero. Any pair of species may be in one out of three relationships:  mutually exclusive ($c_{i,j} = c_{j,i} = \infty$), non-interfering ($c_{i,j} = c_{j,i} = 0$) or dominance  ($c_{i,j} = \infty, \ c_{j,i} = 0$, meaning that $j$ is superior to $i$).  Figure \ref{fig1}b provides a demonstration of these possibilities.

\begin{figure*}
\begin{center}
\includegraphics[width=9cm]{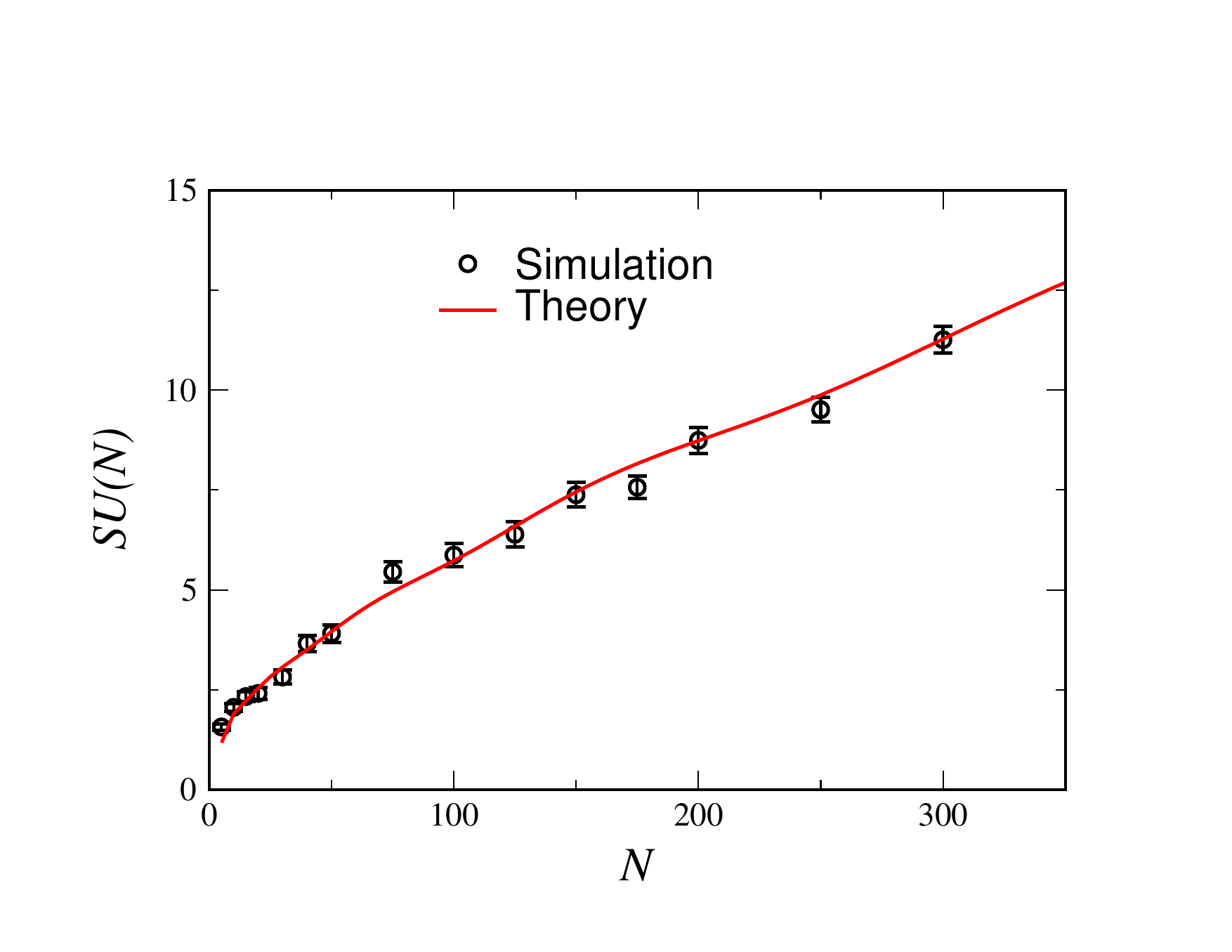}
\end{center}
\caption{The number of SU states as a function of $N$ for the asymmetric zero-infinity model (note the linear scale, in contrast to Fig. \ref{fig2}). Results were obtained from an average over simulations of random networks with ${\tilde p}=0.5$, $N$ running from 5  to 300. Points correspond to the number of SU states in a single realization, full line is the exact sum over $S$ of (\ref{eq4}). The asymptotic relationship (\ref{eq5}) converges to this sum very slowly, see SI section II. } \label{fig3}
\end{figure*}

To interpret this scenario geometrically, one needs two types of links. First, every two non-interfering nodes are connected with undirected links (full lines in Figure \ref{fig1}c), as in the symmetric case. Second, any pair of nodes that admit dominance relationships are connected by a different type of directed link (a dashed line, with an arrow pointing towards the inferior species, in \ref{fig1}c). A stable subcommunity in this case is a clique of non-interacting species as before, but the condition for uninvadablity is different. It is not enough for a clique to be maximal; to be an SU it should fulfill another requirement, namely, that every node not in the clique should be dominated by at least one species in the clique. The expected number of such cliques is then,
\begin{equation}\label{eq4}
SU(N,S) = \binom{N}{S} {\tilde p}^{2\frac{S(S-1)}{2}} (1-{\tilde p}^S)^{N-S}.
\end{equation}
Here ${\tilde p}$ is the chance that  a randomly chosen $c_{i,j}=0$, which leads to  a factor of $2$ in the exponent as compared to the symmetric case where $p$ is the chance that both $c_{i,j}=0$ and $c_{j,i}=0$. The last factor, $(1-{\tilde p}^s)^{N-S}$, reflects the condition that none of the other $N-S$ nodes has a directed dominance links to all the clique member, meaning that all other nodes are inferior with respect to at least one species in the clique.  This seemingly minor modification changes the asymptotic growth mode from superpolynomial to sublinear:
\begin{equation}\label{eq5}
SU(N) \sim \frac{N}{\ln^{3/2}(N)},
\end{equation}
so the conclusion drawn in the symmetric case is relevant, a fortiori,  for asymmetric  communities: there is no exponential growth in the number of $SU$s with $N$, and consequently no niche-neutral transition at large $N$.  The derivation of Eq. (\ref{eq5}) from (\ref{eq4}) is explained in section II of the SI and the agreement between the results and numerical simulations is demonstrated in Figure \ref{fig3}.

Another important result  derived in the SI is an expression for $S^{*}$, the typical number of species in a single SU state, i.e., the maximum of $SU(N,S)$. For both the symmetric model and the asymmetric model we obtain,
\begin{equation}\label{eq6}
S^{*} \approx \frac{\ln(N)}{\ln(1/p)}.
\end{equation}
Accordingly, for $N$ a few hundreds and $p$ values that are not vanishingly small, the typical number of species in a single SU is $5-8$. This estimation agrees with the results of our simulations of the zero/infinity model and  with simulations of the Lotka-Volterra dynamics, Eq. (\ref{eq1}), to which we now turn.

\begin{figure*}
\begin{center}

\includegraphics[width=9cm]{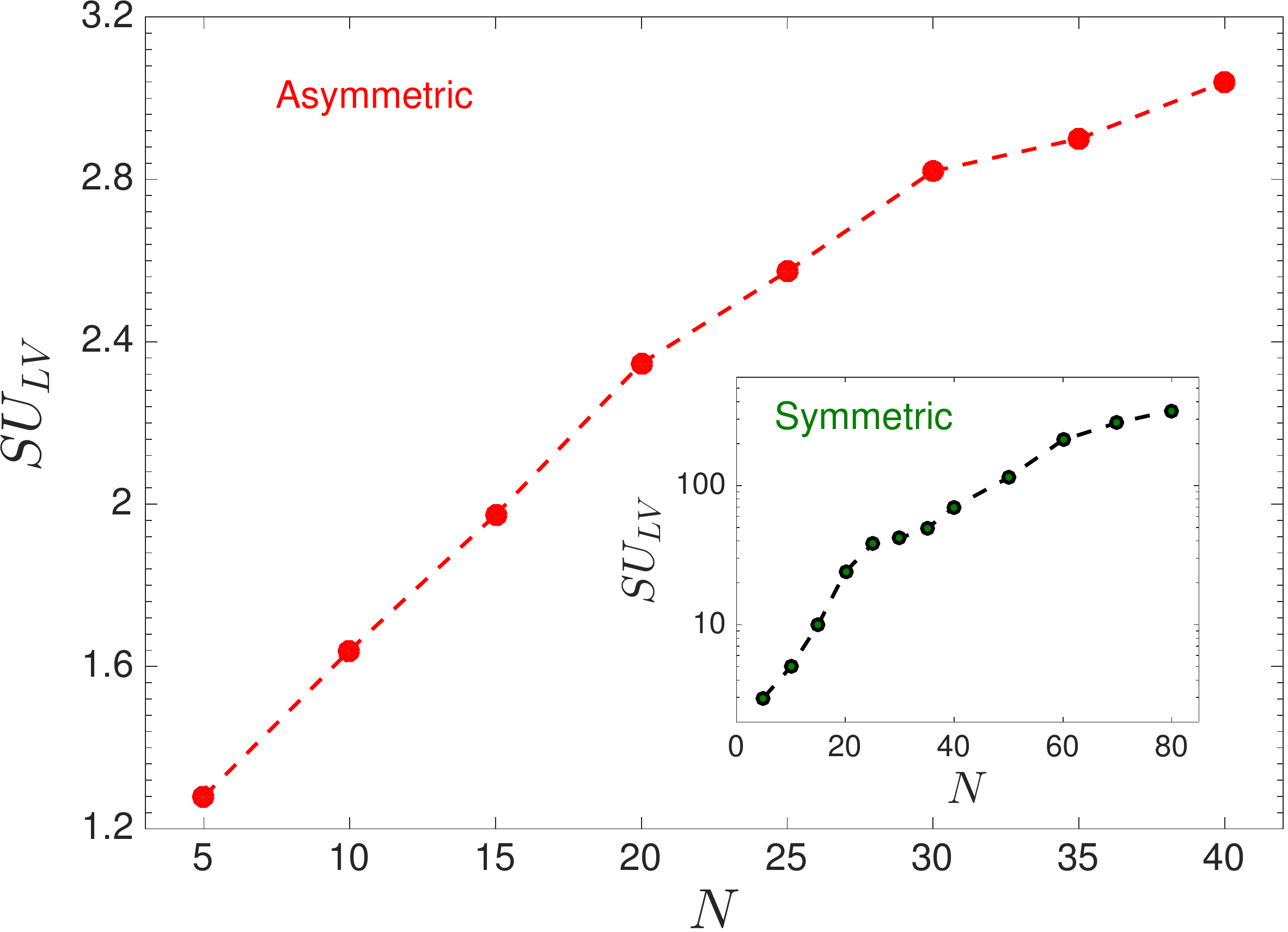}
\end{center}
\caption{$SU_{LV}$ is the average number of SU states for a Lotka-Volterra system (Eq. \ref{eq1}) with continuous $c_{i,j}$ drawn from a Gamma distribution  with ${\bar c} = 1$ and $\sigma^2 = 1$.  For $N \leq 20$ the number of states has been obtained from a comprehensive survey of all $2^N$ possible combinations, while for $N>20$ SU's were identified by integrating Eq. \ref{eq1}, from random initial conditions, until it reaches a SU state, and iterating this scheme  200000 times. In the main panel the results are presented for the asymmetric case, while the inset shows the results for the symmetric case. In both cases  the subexponential growth of the number of $SU_{LV}$ with $N$ is manifested, and the theoretical predictions for the zero-infinity limit  [Eqs. (\ref{eq3}) and (\ref{eq5})] are way above the numbers obtained here (see supplementary). While up to $N=20$ the symmetric case appears to grow exponentially as seen by \cite{gilpin1976multiple}, above this value the graph turns over.   \label{fig5}}
\end{figure*}

The model considered here is an extreme case of a generic competition system, where the interaction matrix may be replaced by a zero-one adjacency matrix. We see no reason to believe that the generic system with finite $c_{i,j}$s falls into a different equivalence class. Actually our numerics for the case where the $c_{i,j}$s are picked from a Gamma distribution indicates a sub-exponential growth in the number of SUs even for symmetric Lotka-Volterra matrices with  finite moments $c_{i,j}$s (see Figure \ref{fig5}), and suggest that the number of SU's obtained in Eqs. (\ref{eq3}) and (\ref{eq5}) is an upper bound for the number of SU's of generic interaction matrices.

 To underscore these statements and to bridge the gap between the zero/infinity limit and the standard generalized Lotka-Volterra system with continuous $c_{i,j}$, an intermediate model is presented in the methods section and analyzed in section III of the SI. In this intermediate model $c_{i,j}$s are either zero or take a finite value $A$, so the competition matrix elements have finite mean and variance, allowing for a fair comparison with the continuous case. This ``binary" model is shown analytically to reproduce exactly the results of the zero/infinity model for sufficiently strong interactions for finite $N$ and generically in the  large $N$ limit. Numerical simulations show that indeed the number of SU's is an upper bound to the continuous Lotka-Volterra system considered in \cite{gilpin1976multiple,fisher2014transition,kessler2015generalized}.

\begin{figure*}
\begin{center}
\includegraphics[width=9cm]{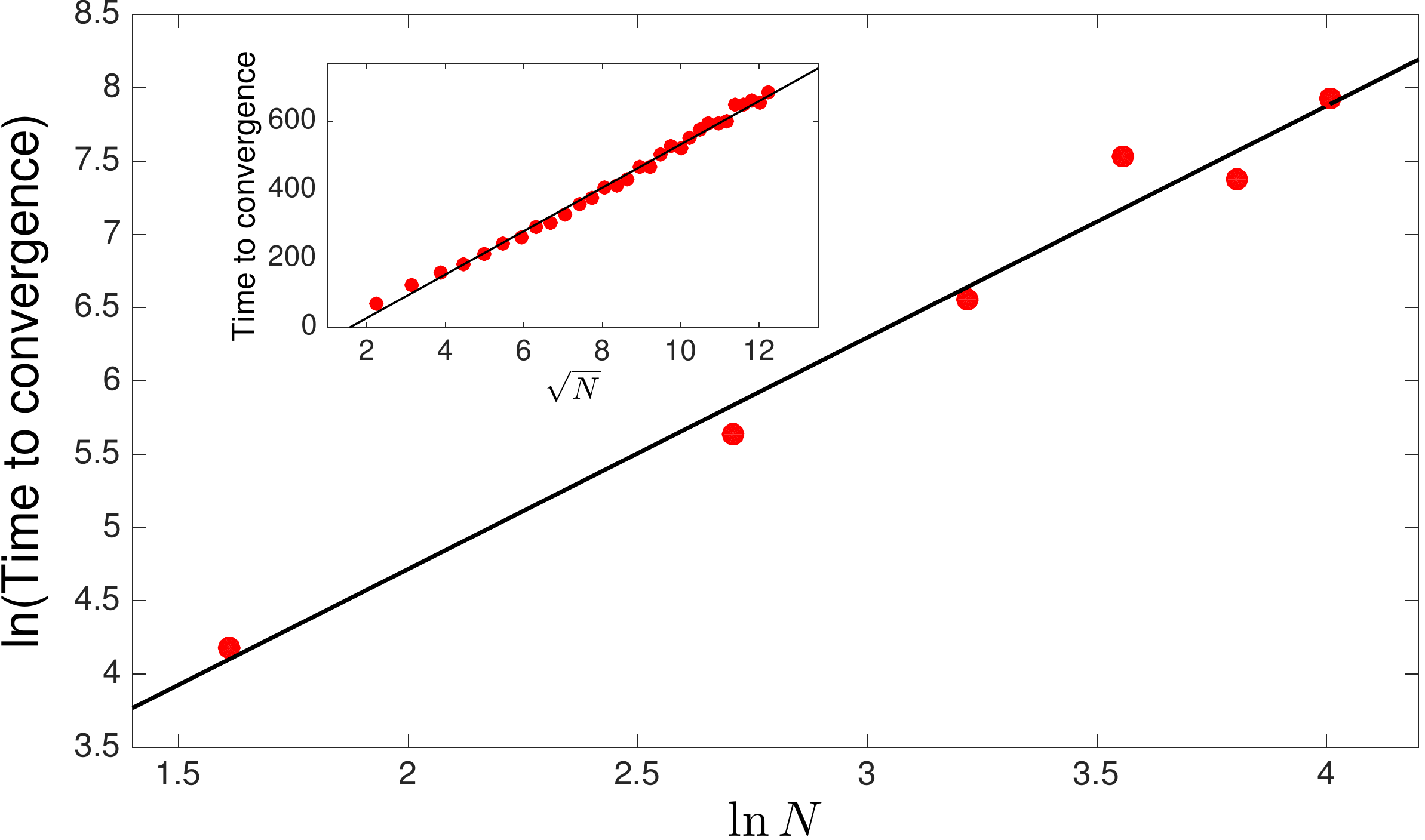}
\end{center}
\caption{ ln(Convergence time) vs. $\ln N$, the species richness of the regional community.  The dynamics of Eq. \ref{eq1} was simulated,  with $c_{i,j}$s that were picked at random from a Gamma distribution with mean $0.9$  and variance $5/N$ (variance should scale with $1/N$ to keep the overall competition stress independent of $N$). Points represent the time it take this system to converge to an SU from random initial conditions. For the asymmetric case (main panel, each point reflects an average over 65 runs each with different random competition matrix) the time to convergence grows like $N^{1.56}$ (thick black line) while for the symmetric case (each point is the average over 2000 runs) the same numerical experiment yields a ${\sqrt{N}}$ dependence.    \label{fig4}}
\end{figure*}

An asymmetric system may support, beyond SU states, other attractive manifolds like limit cycles and chaotic attractors. A simple example is that of a three species system with ``rock-paper-scissors" circular dynamics, species 1 invades 2 but is dominated by 3 and so on. Our simulations indicate that these orbits are rare in the large $N$ limit, and almost all initial conditions eventually arrive at an SU state. However,  the symmetric and the asymmetric systems differ dramatically in the convergence time, as demonstrated in Figure \ref{fig4}.  In symmetric systems the convergence time grows like ${\sqrt{N}}$, while it grows faster than $N^{3/2}$ in the asymmetric case. This feature reflects the presence of many ``almost attractive" orbits and long excursions in the asymmetric system. Accordingly, for a symmetric system one may expect that the dynamics is dominated by long periods in which the system is trapped in a single SU, with ecological regime shifts that drive it from one state to another as suggested in \cite{scheffer2001catastrophic}. In the asymmetric system, on the other hand,   one expects that under the  effect of disturbances (that kick the system out of an SU and send it on a long excursion) the relevance of the SUs becomes negligible  and the system is in the intermittent phase (or, in terms of \cite{fukami2011community}, alternative transient states) that was demonstrated in \cite{kessler2015generalized}.

\section{Discussion}

Community dynamics is the arena on which the evolutionary process unfolds. Darwin's theory of natural selection and the survival of the fittest suggest a mechanism that governs the evolutionary dynamics and the origin of species, but at the same time it makes difficult the task of  explaining the spectacular species richness observed in natural communities. Are there so many ``fittest" species around?

Some researchers believe that the answer for this question is indeed positive: that each of the millions of species observed in nature is superior to its competitors with respect to a certain niche. Others consider this hypothesis as implausible, in particular when the number of resources appear to be small. Moreover, as we pointed out in the Introduction, May's complexity-diversity puzzle implies that a community with substantial niche overlap will collapse, meaning that an almost complete niche-separation is needed to explain the coexistence of any species.

Given that, a lot of attention has been given in the last decades to the patterns observed in local communities, recognized as the elementary building blocks of the system. The dynamics in these communities is affected by local processes such as competition, predation and symbiosis, by migration of species from the regional pool and by stochastic and random effects.

In a recent work~\cite{kessler2015generalized} we have tried to classify the dynamics of a stochastic local community of competing species (in the asymmetric case) along two different axes: the overall strength of competition (average niche overlap, $\overline{c}$) and the fitness differences $\sigma^2$. When $\overline{c} = 0$ all species are non-interacting, as suggested in McArthur-Wilson model, while if $\overline{c} = 1$ and $\sigma = 0$ all \emph{individuals} are equal and the system is described by Hubbell's neutral theory~\cite{maritan1}. Between these two extremes we have found regimes of full and partial coexistence (when $\overline{c}$ is relatively small), regimes of alternate stable states when $\overline{c}$ is large, and in between we observed a region of parameters where the system fails to relax to an SU. Instead, in this regime the dynamics is intermittent, where the community structure changes dramatically over time and the
instantaneous assembly is usually invadable. As noted above, this intermittent phase appears to be related to the long convergence times demonstrated in Fig. \ref{fig4}, combined with the effect of constant perturbations, like the demographic noise in ~\cite{kessler2015generalized}.

A priori, we cannot see a reason to prefer an explicit dynamical model, like the Lotka-Volterra equations (\ref{eq1}), over the zero-infinity dynamics considered in this paper  or its counterpart, the binomial model, presented in the SI. The actual dynamics of interspecific competition is very complex, affected by many factors, and there appears to be no way to map it into a realistic model with  a reasonable number of parameters, not to mention the inference of the values of these parameters from empirical datasets. Both the LV system and the large $\sigma$ limit considered here aim at providing a qualitative picture, explaining the generic characteristics of systems in which species compete, proliferate, migrate and go extinct. We believe that the network analogy is appropriate as long as $\sigma$ is large and  $\overline{c}$ is not too small, which is the most interesting regime.  As explained in the supplementary, the network analogy fails  when $\overline{c} \to 0$ (where all the $N$ species may occupy every local patch, the full/partial  coexistence phase in the language of~\cite{kessler2015generalized}) or when $\sigma \to 0$ (the limit corresponding to the neutral or neutral-like behavior, where the system is governed by noise, see \cite{pigolotti2013species}).  Experimental studies (see, e.g., \cite{carrara2015inferring,mouquet2004immigration}) also suggest that the $c_{i,j}$s are  ${\cal O}(1)$ (i.e., interspecific competition does not differ substantially in magnitude from intraspecific competition) and that the variance is quite large. We believe that the glass transition results of \cite{fisher2014transition} mentioned above, with the assumption that the number of SUs grow exponentially with $N$, are valid in the  opposite limit where both the mean and the variance of the $c_{i,j}$s scale like $1/N$.

 Finally, we believe that the geometric approach presented here may be extended to include more complicated networks. In particular, a foodweb system, like the one considered in~\cite{capitan2009statistical} has a few levels (primary producers, predators, top predators etc.),  within each level different species compete, but the top levels depend on consuming individuals from lower levels. Similarly, in a plant-pollinator and other mutualistic network like those considered in \cite{campbell2011network}, the two networks are interdependent. In both cases the theory of dependent networks~\cite{buldyrev2010catastrophic} may be relevant. The effect of different architectures, like the modular structure found in food webs~\cite{drossel2001influence} and the nested structure associated with mutualistic networks~\cite{bascompte2003nested,suweis2013emergence} is also an interesting factor, and one would like to study this is the context of network dynamics. We hope to address these topics in subsequent publications.

\section{Methods}
  Our aim is to find the number of SU states for a community of competing species described by the generalized Lotka-Volterra equation [\ref{eq1}]. Commonly the $c_{i,j}$s are drawn independently from a uniform, positive semi-definite, distribution with a given mean and variance \cite{fisher2014transition,kessler2015generalized}. A $c_{i,j}$ matrix (for simplicity the examples are given for the symmetric case) may look like,
 \[ C \left( \begin{array}{cccc}
0 & 0.95 & 1.63 & 0.96 \\
0.95 & 0 & 0.48 & 0.97 \\
1.63 & 0.48 & 0 & 1.12 \\
0.96 & 0.97 & 1.12 &0  \end{array} \right).\]
Since we used a Gamma distribution in our simulations, we denote this as the Gamma model.

The mapping of this model to the maximum clique problem (which is exact in the limit where all matrix elements are either zero or infinite) involves two steps of simplification. First we assume that all the elements of the $c_{i,j}$ matrix either are strictly zero or are equal to a finite constant $C \dot A$, so the interaction matrix takes a form, say,
  \[ C \left( \begin{array}{cccc}
0 & A & 0 & A \\
A & 0 & 0 & A \\
0 & 0 & 0 & 0 \\
A & A & 0 &0   \end{array} \right),\]
where we introduced two constants, $C$ and $A$, to distinguish between the overall strength of the competition and the matrix structure. In the SI (section III) we show how these constants should scale such that this "binary" model and the continuum Gamma model matrix elements will have the same mean and variance.

In the binary model every maximal clique is (trivially) stable, since the strength of  competition between all species in such a clique is zero. However, if $A$ is too small such a maximal clique may be invadable by one of the the species outside the clique. In section III of the SI we show that this cannot happen if $C$ is large enough (the condition is $C>1-p$, where $p$ is the fraction of zeros in the competition matrix). Above this critical $C$ our maximum cliques are both stable and uninvadable, and  one may replace the zero-$A$ interaction matrix by the corresponding zero/infinity matrix,
 \[ \left( \begin{array}{cccc}
0 & \infty & 0 & \infty \\
\infty & 0 & 0 & \infty \\
0 & 0 & 0 & 0 \\
\infty & \infty & 0 &0   \end{array} \right).\]

Moreover, we show (in SI III) that this substitution of the infinite-zero model instead of the binomial model becomes \textit{exact} for \textit{any} $C$ in the large-$N$ limit, and provide numerical evidence to show that the number of cliques of this binomial model is an upper bound for the continuum competition (Gamma) model with the same mean and variance.

\begin{acknowledgments}
We thank Lewi Stone for fruitful discussions.  N.M.S.  acknowledges the support of the Israel
Science Foundation, grant no. $1427/15$. D.A.K. acknowledges the support of the Israel Scientific Foundation, grant no. $376/12$
\end{acknowledgments}

\bibliography{references}

\bibliographystyle{pnas2011}

\newpage
\onecolumngrid

\pagebreak
\widetext
\begin{center}
\Large{\bf{Supplementary information to: Communities as cliques}}
\end{center}
\setcounter{equation}{0}
\setcounter{section}{0}
\setcounter{figure}{0}
\setcounter{table}{0}
\setcounter{page}{1}
\makeatletter
\renewcommand{\theequation}{S\arabic{equation}}
\renewcommand{\thefigure}{S\arabic{figure}}

\vspace{1cm}

This supplementary material consists of three sections. In the first two sections we provide the mathematical details of the derivation of Eq. (3) of the main text from Eq. (2), and the derivation of Eq. (5) from Eq. (4), respectively. The aim of the third section is to analyze an intermediate model presented in the method section of the main text.  This model bridges between our infinite $\sigma$ model and the standard generalized Lotka-Volterra system (Eq. (1) of the main text with $c_{i,j}$ that are drawn from a continuous distribution with finite moments).

\section{Asymptotics of $SU(N)$, Symmetric Network}

Let $p$ be the probability of a symmetric coexistence link ($c_{i,j}=c_{j,i}=0$).  Then the number of maximal cliques of size $S$ in a random graph of $N$ nodes is given in Eq. (2) of the main text,  following \cite{bollobas1976cliques}:
\begin{equation} \label{eqa1}
SU(N,S) = {N \choose S} p^{S(S-1)/2} (1-p^S)^{N-S}
\end{equation}
To get the large $N$ asymptotic of this sum we define
\begin{equation}
\alpha\equiv \ln(1/p).
\end{equation}
Using Stirling's formula we can approximate (\ref{eqa1}) for large $N$, $S\ll N$, as
\begin{equation}
\ln SU(N,S) \approx S\ln N - (S+1/2)\ln(S) + S - \frac{1}{2}\ln(2\pi) - \alpha S(S-1)/2 - (N-S)e^{-\alpha S} .
\label{lnC}
\end{equation}
 Taking a derivative with respect to  $S$, we get an equation for $S^*$, the value of $S$ that gives the maximal contribution to the sum in Eq. (\ref{eqa1}):
\begin{align}
0 &= \ln N - \ln S^* - 1/(2S^*) - \alpha(S^*-1/2) + \tilde p^{-S^*} + \alpha e^{-\alpha S^*}(N-S^*)  \nonumber\\
&=\alpha Ne^{-\alpha S^*} + \ln (Ne^{-\alpha S^*}) - \ln S^* - 1/(2S^*) + \frac{1}{2}\alpha + e^{-\alpha S^*} - \alpha S^*e^{-\alpha S^*}
\end{align}
The dominant balance is between the terms $\alpha Ne^{-\alpha S^*}$ and  $-\ln S^*$, giving
\begin{equation}
\alpha Ne^{-\alpha S^*} = \ln S^*
\end{equation}
so that
\begin{equation}
S^* \approx \frac{\ln N}{\alpha} + {\cal{O}}(\ln(\ln(\ln(N)))).
\end{equation}
As a first approximation one may assume that the logarithm of the sum (\ref{eqa1}) is equal to the contribution from $S^*$ alone.  Plugging this into Eq. (\ref{lnC}) one finds,
\begin{equation} \label{eqa2}
\ln SU(N) \approx \ln SU(N,S^*)\approx \frac{\ln^2 N}{2\alpha} - \frac{\ln N}{\alpha}\ln\left(\frac{\ln N}{e\alpha}\right).
\end{equation}
In principle, we have to include also the contribution from a Gaussian integral in $S-S^*$ around $S^*$.  However, the coefficient of $(S-S^*)^2$ in this integral is given by,
\begin{equation}
\frac{d^2}{dS^2} \ln SU(N,r)\Big |_{r=S^*} \approx -N\alpha^2 e^{-\alpha S^*} \approx \ln S^* \alpha \approx \alpha\ln \left(\frac{\ln N}{\alpha}\right)
\end{equation}
so in fact the correction decays so quickly around $S^*$ that only the leading term (\ref{eqa2}) contributes (note that the sum is discrete). This leads to Eq. (3) of the main text.

\section{Asymptotics of $SU(N)$, Asymmetric Network}
\begin{figure}[h]
\vspace{-1cm}
\begin{center}
\includegraphics[width=9cm]{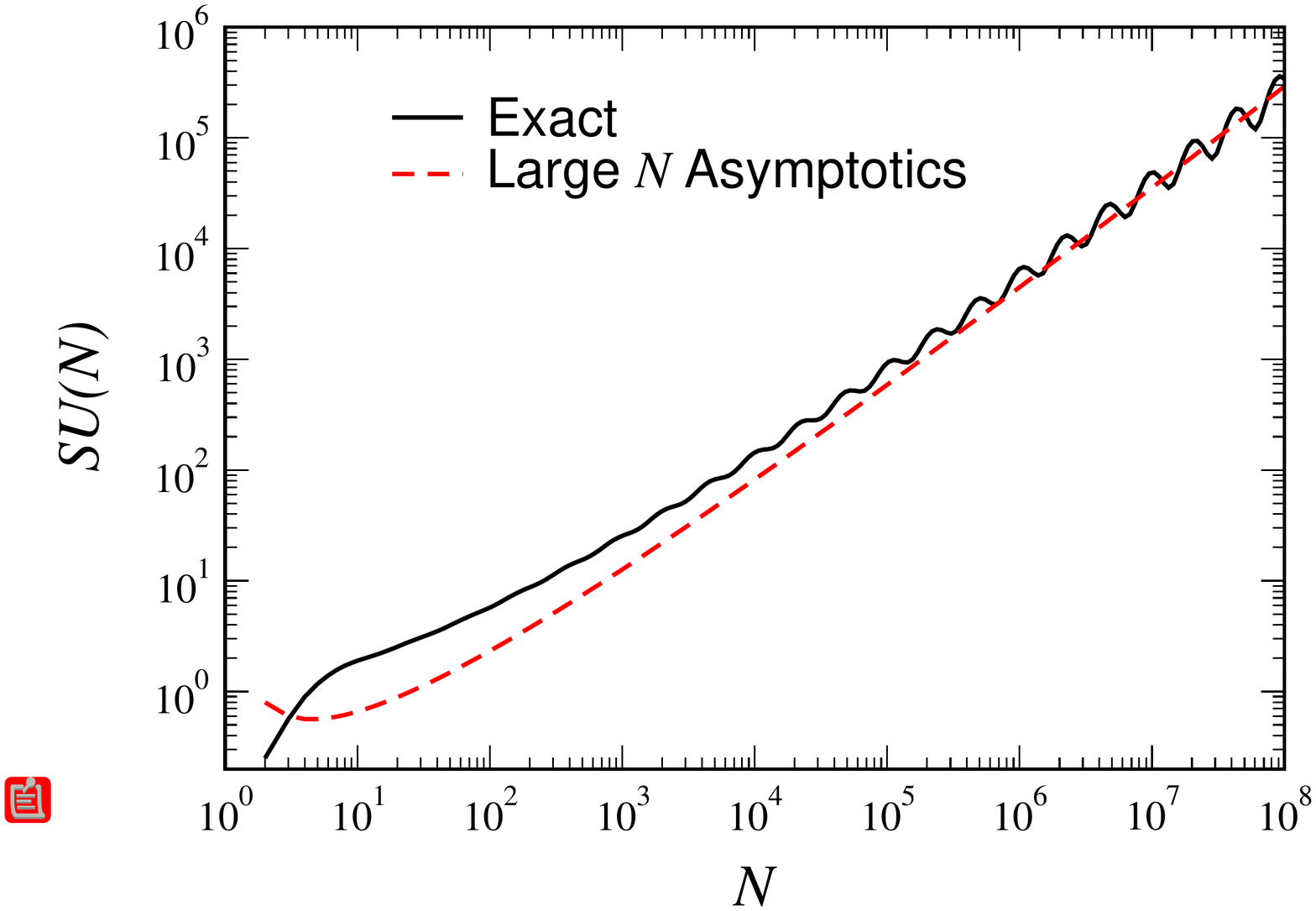}
\end{center}
\vspace{-1cm}
\caption{The number of SUs in the asymmetric case: a comparison between the exact sum over $S$ in Eq. (\ref{eqx}) (black solid line) and the asymptotic formula (\ref{assyp}) (dashed red line).    \label{figsum}}
\end{figure}

\noindent For asymmetric links, we found that (Eq. (4) of the main text)
\begin{equation} \label{eqx}
SU(N,S) = {N \choose S} \tilde{p}^{S(S-1)} (1-\tilde{p}^S)^{N-S}
\end{equation}
Again, writing things in terms of $\alpha\equiv \ln(1/\tilde{p})$, we have
\begin{equation}
\ln SU(N,S) \approx S\ln N - (S+1/2)\ln(S) + S - \frac{1}{2}\ln(2\pi) - \alpha S(S-1) - (N-S)e^{-\alpha S}
\label{lnCA}
\end{equation}
The dominant $S$, $S^*$, obeys
\begin{align}
0 &= \ln N - \ln S^* - 1/(2S^*) - \alpha(2S^*-1) + \tilde p^{-S^*} + \alpha e^{-\alpha S^*}(N-S^*)  \nonumber\\
&=\alpha Ne^{-\alpha S^*} + \ln (Ne^{-\alpha S^*}) -\alpha S^* - \ln S^* - 1/(2S^*) + \alpha + e^{-\alpha S^*} - \alpha S^*e^{-\alpha S^*}
\end{align}
The dominant balance is now between the terms $\alpha Ne^{-\alpha S^*}$ and  $-\alpha S^*$, giving
\begin{equation}
S^* \approx \frac{\ln N}{\alpha} - \frac{\ln\left(\frac{\ln N}{\alpha}\right)}{\alpha}
\end{equation}
Using this, the ${\cal{O}}(\ln^2 N)$ terms cancel, leaving us with
\begin{equation} \label{assyp}
\ln SU(N) \approx SU(N,S^*)\approx \ln N - \frac{3}{2}\ln\left(\frac{\ln N}{\alpha}\right) - \frac{1}{2}\ln{2\pi},
\end{equation}
and this yields the result that appears in Eq. (5) of the main text. Note that the convergence of this expression to the exact sum over $S$ in Eq. (\ref{eqx}) is quite slow and nonmonotonic, see Figure \ref{figsum}.

Again, the decay of $SU(N,S)$ is so rapid,
\begin{equation}
\frac{d^2}{dS^2} \ln SU(N,r)\Big |_{r=S^*} \approx -N\alpha^2 e^{-\alpha S^*} \approx - \alpha S^* \approx -\frac{\ln N}{\alpha}
\end{equation}
that only the $S^*$ term contributes to the sum asymptotically.

\newpage

\section{The Binary Model}

In the main text we discussed the connection between SUs and cliques in the limit where the competition matrix elements are either zero or infinite, so two species $i$ and $j$ may be noninteracting ($c_{ij} = c_{ji}=0$), mutually exclusive ($c_{ij} = \infty, \  c_{ji}= \infty$) or in dominance relationships ($c_{ij} = \infty, \  c_{ji}=0 \ {\rm or} \ c_{ij} = 0, \  c_{ji}=\infty$).   This presentation leads us immediately to the notion of cliques and to the conclusions we drow from the equivalence between cliques and SUs. Here we would like to discuss the relationships between this extreme limit and the ``standard" GLV description of the system, in which the $c_{ij}$s are picked at random from a continuous distribution with finite moments. To be specific we will use, as in \cite{kessler2015generalized},  Gamma distributed $c_{i,j}$s and denote this version of the generalized Lotka-Volterra as the Gamma model.

To bridge the gap between the zero-infinite  and the Gamma model, we present here an intermediate scenario, the binary model, which allows us to obtain a few analytic results that establish the relevance of the clique-based analysis. At the same time, this model, which has finite $\sigma$ and $C$, facilitates a numerical comparison with the standard GLV model. These three types of interaction matrices - Gamma, binary and zero-infinite -  are illustrated in Figure \ref{figS1}. For simplicity our discussion is presented for the symmetric case, but the results are general.

\begin{figure} [h] \label{x}
\[ \begin{array}{ccc}
 {\rm Gamma}&  {\rm Binary} & \textit{zero/infinity} \\ \\
  \left( \begin{array}{cccc}
0 & 0.95 & 1.63 & 0.96 \\
0.95 & 0 & 0.48 & 0.97 \\
1.63 & 0.48 & 0 & 1.12 \\
0.96 & 0.97 & 1.12 &0  \end{array} \right)\quad %
& \quad C  \left(  \begin{array}{cccc}
0 & A & 0 & A \\
A & 0 & 0 & A \\
0 & 0 & 0 & 0 \\
A & A & 0 &0  \end{array} \right) \quad%
& \quad \left( \begin{array}{cccc}
0 & \infty & 0 & \infty \\
\infty & 0 & 0 & \infty \\
0 & 0 & 0 & 0 \\
\infty & \infty & 0 &0  \end{array} \right)
\end{array}
\]
\caption{Three types of interaction matrices $c_{i,j}$. The left matrix corresponds to a four species community in which the niche overlap between species is assumed to be a random number taken from a Gamma distribution. To the right one sees an interaction matrix that corresponds to the zero/infinity limit considered in the main text, where every two species are either non-interacting or are mutually exclusive. The interaction matrix in the middle ($C$ and $A$ are constants, see below) exemplifies our binary model: like the infinite $\sigma$ case is has only two types of interactions, but unlike it, it admits a finite value for $\sigma$, allowing for a comparison with GLV systems with continuous $c_{i,j}$s that have the same parameters.  For simplicity we present examples for the symmetric case, where the level of competition between every two species is characterized by a single number that corresponds to the niche overlap between these two species. In the non-symmetric case $c_{ij} \neq c_{ji}$, but all  other features of the three models are the same. \label{figS1}}
\end{figure}

 To begin, let us rewrite the GLV equation used in the main text,
 \begin{equation} \label{eqS1}
\frac{dx_i}{dt} = x_i - x_i \left(x_i + C \sum\limits_{j \neq i }^N c_{i,j} x_j \right).
\end{equation}
 where now (following \cite{kessler2015generalized}) we assume that the $c_{i,j}$s are normalized such that their average is unity and $C$ sets the overall scale of the interaction. In the binary model, the $c_{i,j}$s are either $A$ or zero, so to fix their average to unity we must choose  $A=1/(1-p)$, where $p$ is the probability of $c_{i,j}=c_{j,i}=0$.
  It follows that the variance of the $c_{i,j}$s is
\begin{equation}
\textrm{Var}[c_{i,j}] = \sigma^2 = A^2 p (1-p) = \frac{p}{1-p}.
\end{equation}
Thus, our binary model allows us to control, as in the standard Gamma distribution model, both the overall competition strength $C$ and the variance of the competition, $\sigma^2$, and to compare Gamma and binary competition matrices with the same mean and variance.

The binary model is similar in many respects to the infinite $\sigma$ model discussed in the main text. Indeed, its maximal cliques are {\em precisely} the SUs for $C$ greater than some threshold value $C_t<1$.  It is trivial to see that a maximal clique is a solution of the GLV equation (\ref{eqS1}), since for that subset of species the model is noninteracting, and the state with $x_i = 1$ for all species $i$ represented in the maximal clique is feasible and  stable.  The nontrivial task is to find under what conditions the solution is also uninvadable, rendering it an SU.  The equation of invadability for an absent species $\alpha$ is
\begin{equation}
\dot{x}_\alpha = x_\alpha\left[1 - C\sum \limits_{i \in \boldsymbol{\mathcal{S}}} c_{\alpha,i} \right],
\end{equation}
where the sum is over all present species $i \in {\rm clique}$. In the binary model $c_{i,j}$ is either $A$ or zero, so $\sum \limits_{i \in \boldsymbol{\mathcal{S}}} c_{\alpha,i}$ is simply $A$ times the integer $m_\alpha^{\boldsymbol{\mathcal{S}}}$,  the number of species (in the clique $\boldsymbol{\mathcal{S}}$) with nonzero $c_{\alpha,i}$ element with the spececies $\alpha$.  Thus, the solution is uninvadable (and so is an SU) as long as,
\begin{equation} \label{eqS5}
C < \max_\alpha \frac{1}{\sum \limits_{i \in \boldsymbol{\mathcal{S}}} c_{\alpha,i}}  \qquad {\rm or} \qquad C < \max_\alpha \frac{1}{A m_\alpha^{\boldsymbol{\mathcal{S}}} }
\end{equation}
the maximum is taken from the values for all the species $\alpha$ which are \emph{not} in the clique. Since any species out of the clique has at least one enemy in the clique (otherwise, the clique is not maximal) for any $C > 1-p$ no clique is invadable.

For example, let us consider the following competition matrix:
 \[ C \left( \begin{array}{cccc}
0 & 0 & 0 & 6 \\
0 & 0 & 0 & 0 \\
0 & 0 & 0 & 0 \\
6 & 0 & 0 &0   \end{array} \right).\]
This is a 4-species realization of the binary model, and since there are 6 possible competition terms from which only one is active, it follows that $p=5/6$ hence $A = 6$.

Clearly, species $\{1,2,3\}$ constitute a maximum clique: they all do not compete with each other, while species $\{4\}$ suffers from competition with $\{1\}$. Eq. \ref{eqS1} for $x_4$ (when the island is occupied by species $\{1,2,3\}$, each with abundance one)  now reads,

\begin{equation}
\frac{dx_4}{dt} = x_4 - x_4^2 - 6 C x_1 x_4 = (1- 6C)x_4 - x_4^2,
\end{equation}
so $x_4$ may invade (its linear growth term is positive) only if $C<1/6$, but above this value the maximal clique is indeed an SU. This agrees with Eq. (\ref{eqS5}) as $A=6$ and $m_4^{\boldsymbol{\mathcal{S}}} = 1$.

\begin{figure}[h]
\begin{center}
\includegraphics[width=9cm]{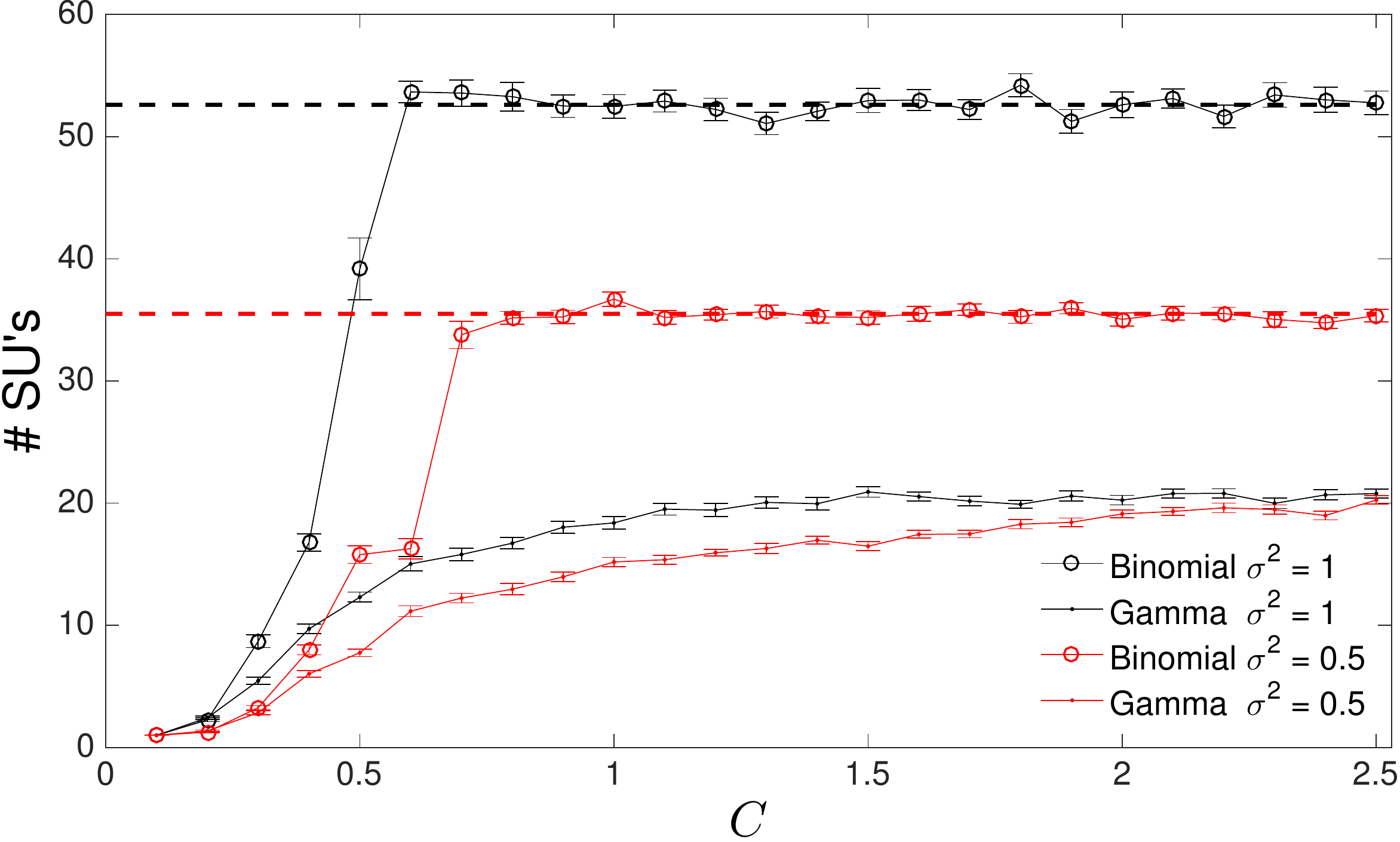}
\end{center}
\caption{The number of SU states as a function of $C$ for the binary symmetric model and for the Gamma model, both with $N=20$. Graphs show the average and the standard deviation (error bars) of the number of SUs for  random matrices with $\sigma^2 = 0.5$ and $\sigma^2 = 1$. For the binary model one can clearly see the saturation above some $C_t$, where all the maximal cliques are SUs and further increase in the value of $C$ has no effect. The value of $C_t$ decreases as $\sigma^2$ (and $p$) increases, in agreement with our predictions. The dashed lines are the expected number of maximal cliques, computed  from Eq. (2) of the main text with the relevant values of $p$ ($1/2$ and $1/3$), and thus are the number of SUs in the infinite-$\sigma$ model with the same value of $p$. Clearly, the number of SU saturates to these values, which are way above $N$. For the Gamma model, on the other hand, the maximum number of SUs is about $N$, and one can see that the number of SUs in the binary model is always an upper bound to the number of SUs in the Gamma model. Results were obtained by averaging over 65 different matrices in each case.  \label{figS2}}
\end{figure}

\begin{figure}[h]
\begin{center}
\includegraphics[width=9cm]{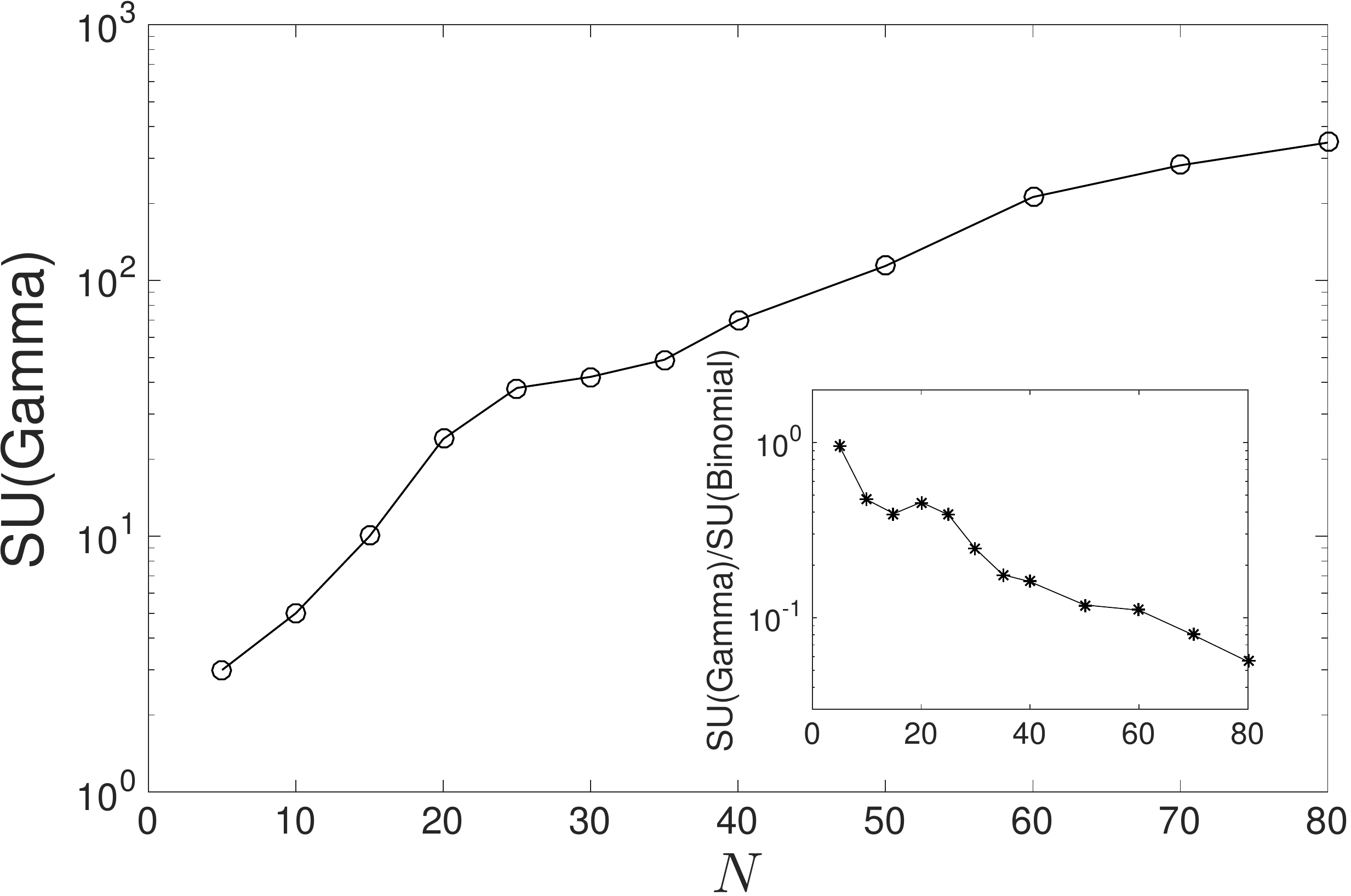}
\end{center}
\caption{ The number of SUs for the Gamma distributed, symmetric Lotka-Volterra model,  as shown in Fig. 4 of the main text, is reproduced in the main panel. The inset shows these numbers divided by the number of cliques in the binary model with the same values of $C=1$ and $\sigma$. One realizes that the number of SUs obtained for the binary model provides an upper bound for the Gamma distributed $c_{i,j}$ case. Since for this value of $C$ the number of SUs in the binary model is precisely the number of maximal cliques of the infinite-$\sigma$ model, which we have shown to be subexponential, this is a clear indication that the Gamma distributed Lotka-Volterra model behaves likewise.  \label{figS3}}
\end{figure}

Within the binary model, the general features of the Gamma distribution model are preserved.  For example, if $C$ is smaller than some threshold value $C<C_\textrm{1}$ (using the notations defined in fig 1 of \cite{kessler2015generalized}) there is exactly one SU, namely the state with all species present.  This is clearly not a maximal clique for general $p$. Then, up to a second critical point $C_2$, there is still only a single SU, with however, some  species are missing. The general trend is that below $C_t$ the number of SUs falls below the number of maximal cliques, and decreases monotonically as $C$ decreases.  This is demonstrated in Fig. \ref{figS2}, where we have plotted the number of SU's in the binary model (averaged over a number of realizations) as a function of $C$ for $N=20$ and two different values of $\sigma$.

Together with the results of the binary model we have  plotted the average number of SUs in the Gamma distribution model for the same values of $N$, $C$ and $\sigma$. One sees that the number of SUs in the Gamma distribution model is significantly smaller. In the limit $C \to \infty$ every pair of species in the Gamma model is mutually exclusive so every species is an SU, meaning that the number of SUs in the Gamma model approaches $N$ for large $C$.  This numerical evidence, together with those presented in Fig. 4 of the main text and in Fig. \ref{figS3} here,  reinforce our claim that the number of SUs in the Gamma distribution case also grows slower than exponentially with $N$.

\end{document}